# Enhancement of the high-field critical current density of superconducting MgB$_2$ by proton irradiation

Y. Bugoslavsky*†, L.F. Cohen*, G.K. Perkins*, M. Polichetti*‡, T.J.Tate*, R. Gwilliam§ and A.D. Caplin*

* *Centre for High Temperature Superconductivity, Blackett Laboratory, Imperial College, London SW7 2BZ, UK*

§ *EPSRC Ion Beam Centre, University of Surrey, Guildford, Surrey GU2 7RX, UK*

† On leave of absence from General Physics Institute, Moscow, Russia

‡ Visiting from INFM – Dipartimento di Fisica, Universita' di Salerno, Salerno, I-84081, Italy

**A relatively high critical temperature, $T_c$, approaching 40 K, places the recently-discovered [1] superconductor magnesium diboride (MgB$_2$) intermediate between the families of low- and copper-oxide-based high-temperature superconductors (HTS). Supercurrent flow in MgB$_2$ is unhindered by grain boundaries [2, 3], unlike the HTS materials. Thus, long polycrystalline MgB$_2$ conductors may be easier to fabricate, and so could fill a potentially important niche of applications in the 20 to 30 K temperature range. However, one disadvantage of MgB$_2$ is that in bulk material the critical current density, $J_c$, appears to drop more rapidly with increasing magnetic field than it does in the HTS phases [4]. The magnitude and field dependence of $J_c$ are related to the presence of structural defects that can "pin" the quantised magnetic vortices that permeate the material, and prevent them from moving under the action of the Lorentz force. Vortex studies [3] suggest that it is the paucity of suitable defects in MgB$_2$ that causes the rapid decay of $J_c$ with field. Here we show that modest levels of atomic disorder, induced by proton irradiation, enhance the pinning, and so increase $J_c$ significantly at high fields. We anticipate that chemical doping or mechanical processing should be capable of generating similar levels of disorder, and so achieve technologically-attractive performance in MgB$_2$ by economically-viable routes.**

A Type II superconductor undergoes the transition to the normal state at the upper critical field, $H_{c2}$. However, the ability to carry dissipation-free current ceases at a lower field, the irreversibility field, $H_{irr}$. Above $H_{irr}$ the Lorentz force on the vortices is large enough for them to become detached from pinning defects, and virtually free to move. In MgB$_2$ powder (and also wires and tapes), $H_{irr}$ is approximately half of $H_{c2}$ [5], so that there is an extended field domain where there might be a useful $J_c$ if only the pinning could be strengthened.

Ion irradiation is a reproducible means of inducing crystalline disorder. Energetic ions displace atoms from their equilibrium lattice sites, creating a variety of defects, including vacancies and interstitials. Such defects tend to depress the superconducting order parameter locally, and thereby create pinning sites for the vortices. We chose protons for this initial study because at the maximum beam energy available to us of 2 MeV, they can penetrate 40 to 50μm into MgB$_2$. A series of irradiations were performed in order to create significant defect densities. The probability of displacement per atomic site (dpa) can be simulated using commercial software [6], and we aimed to generate as uniform a profile as possible through the sample depth (Fig.1).

As in our previous study [3], we selected fragments of about 100 μm size from commercial MgB$_2$ powder (Alfa Aesar Co., 98% purity). Several samples were prepared, each of 20 fragments embedded in silver-loaded epoxy and then polished, so that a defined geometry was obtained, with an estimated thickness of the MgB$_2$ fragments of about 50 microns. Also, this configuration ensured that good electrical and thermal contact with the conducting substrate was maintained during irradiation. Two samples were irradiated to fairly uniform damage levels of ~1% and 5% dpa respectively, and one sample was more lightly damaged (~0.04% dpa through most of its volume), but in a less-uniform manner (see Figure 1 legend). After the irradiation, the fragments were extracted from the epoxy and their magnetic moments measured.

The critical temperature, $T_c$, drops with irradiation (Figure 2). This could be caused by a number of possible factors, e.g., depression of the density of states at the Fermi-level by the disordering. Also, the transition broadens considerably in the irradiated samples; the inhomogeneity of the damage must contribute significantly to the breadth. It should be noted that in the 1 and 5% dpa samples, the onset of the transition is lowered too, i.e. there is no indication of the presence of any residual undamaged material in these samples.

We obtain the critical current density $J_C$ from measurements of the irreversible magnetisation in the standard fashion [7]. Irradiation does degrade $J_C$ at low fields (Figure 3), in part at least because of the reduction in $T_c$. However, the crucial result is that the field dependence of $J_C$ flattens significantly in the irradiated samples. Even the lowest level of damage reduces the downward slope of the $\log(J_C)$-H curve. Higher irradiation doses further decrease this slope by a factor of ~2. The combination of low-field degradation and high field improvement yields a cross-over, so that, for example, the 1% dpa sample at 20 K has a higher $J_C$ than the virgin sample at fields above about 2.5 T.

Another way of quantifying the field-dependence of $J_C$ is by defining the irreversibility field, $H_{irr}$, above which $J_C$ drops below a useful level (Figure 4). We show also the upper critical field, $H_{c2}$, for non-irradiated $MgB_2$, as measured by Bud'ko et al. [5]. Irradiation progressively increases the slope, $dH_{irr}/dT$, and for the 5% dpa sample it closely approaches the virgin sample $dH_{c2}/dT$. Very heavily damaged (by neutron irradiation) $MgB_2$ has been studied by Karkin et al. [8], who found that the disorder strongly reduces $T_c$, but leaves $dH_{c2}/dT$ almost unchanged; they did not attempt to measure $J_C$. It is therefore likely $dH_{c2}/dT$ is unchanged in our samples too.

The one other report of an enhancement of $H_{irr}$ in $MgB_2$ is in thin (sub-micron) films deposited by laser ablation [9]. There too there is a depression of $T_c$, and the slope of $dH_{irr}/dT$ is even higher than in our irradiated samples (Figure 4). These films are nano-crystalline, and it seems likely that in them the grain boundaries are responsible for the strong vortex pinning.

At this stage, the detailed nature of the defects in irradiated $MgB_2$ is uncertain. The incident protons should be able to cause displacements on both Mg and B sublattices, and interstitials may form loops between the planes. At the end of their track, the protons may simply diffuse out, form $MgH_2$, or react to form boranes. In any event, the defects form stronger pinning sites than those that are native in $MgB_2$, as shown by the sharp reduction in slope of the $J_C$ versus $H$ plots.

Proton irradiation is clearly impractical for large-scale manufacture of $MgB_2$ high current conductors, but these results demonstrate that modest disorder - at a level that should be achievable by chemical doping or alloying - in $MgB_2$ bulk can significantly improve $J_C$ at high magnetic fields. The defect species and concentration will need to be optimised so as to minimise the reduction of $T_c$, while maximising the enhancement of $H_{irr}$.

The performance of HTS conductors has improved steadily over their 10 years of development [4]. The best commercially-fabricated long-length tapes incorporating the BiSrCaCuO2223 phase achieve $J_C$'s of about $1.5 \ 10^5$ A cm$^{-2}$ at 20 K and zero applied field, dropping by a factor of about 3 in a field of 2 T. Both this absolute value of $J_C$ and its (approximately exponential) field-dependence are very close to those of irradiated $MgB_2$ (Figure 3).

We suggest that at temperatures around 20 K, which can be achieved readily with standard cryocoolers, and at fields up to a few Tesla, $MgB_2$ conductors with achievable pinning enhancement will be competitive in performance with HTS conductors. The latter are costly to produce, so that if $MgB_2$ conductors can be fabricated cheaply - the constituents themselves are inexpensive - several applications of superconductivity, e.g. for open-access medical Magnetic Resonance Imaging magnets [10], will become much more economically-attractive.

Acknowledgements:
We are grateful to Prof. Marshall Stoneham and Dr Juan Matthews for helpful advice on radiation damage. This work was supported by the UK Engineering and Physical Sciences Research Council.

**Correspondence should be addressed to Y.B. (e-mail y.bugoslav@ic.ac.uk)**

**Figure captions**

Figure 1. Simulated [6] damage profiles produced in $MgB_2$ by proton irradiation. The damage is measured in units of displacements per atom (dpa), but for protons at these energies the absolute values are known only to within a factor of about 2. A monoenergetic proton beam causes an almost flat distribution of damage for most of its range, and then a sharp peak towards the end of the track. The broken line shows simulated damage for such a beam of energy 2 MeV and a fluence of $10^{16}$ cm$^{-2}$, as was used to irradiate the sample denoted 0.04% dpa. Inset: the same data on an expanded scale. In order to create fairly uniform damage through the depth of a second sample, 15 consecutive implants were made with the beam energy varied between 2 MeV and 400 keV (solid line); the average damage is 1% dpa. A third sample was irradiated similarly, but at 5 times higher fluence, yielding 5% dpa damage. In order to minimise sample heating, the beam current was maintained at 2μA/cm$^2$.

Figure 2. Effect of irradiation on the superconducting transition. As the damage increases, the onset critical temperature, $T_c$, decreases and the superconducting transition broadens, reflecting the disorder. The samples were zero-field cooled, a field of 10 mT applied, and the magnetic moment $m$ measured while warming.

Figure 3. Effect of irradiation on the field-dependence of $J_C$ at 20 K; the behaviour at other temperatures is similar. $J_C$ is obtained from the magnetisation hysteresis width, using the Bean model.[7] Main panel: $J_C(H)$ of the virgin and irradiated samples at 20 K, with each sample normalised to its zero field value $J_C(0)$. Slower depression of $J_C$ by the field, as occurs even with the lowest dose, signifies more efficient vortex pinning in the irradiated samples. At higher doses, the improvement in the gradient of log($J_C$)-H curves saturates, but $J_C(0)$ is suppressed, therefore there is an optimal level of disorder for enhancement of $J_C$ at high fields. Inset: absolute values of $J_C$, which are somewhat uncertain, because the radiation damage is non-uniform. Also, in these polycrystalline fragments, it is possible that defects have migrated to grain boundaries, interrupting the circulation of supercurrent.

Figure 4. Temperature of the irreversibility field $H_{irr}$ for different damage levels. The irreversibility field is that at which $J_C$ becomes immeasurably small; here we use a criterion of 1 kA/cm$^2$. There is a systematic increase of the slope of $H_{irr}(T)$ curve with increasing dose. For the 5% dpa sample, this slope closely approaches the slope of $H_{c2}(T)$ in the virgin sample (filled squares). The irreversibility field of the nanocrystalline $MgB_2$ thin films reported by Eom et al. [9] is shown too.

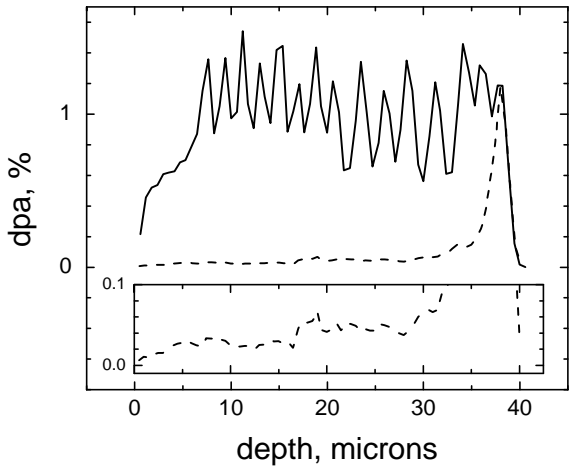

Fig. 1

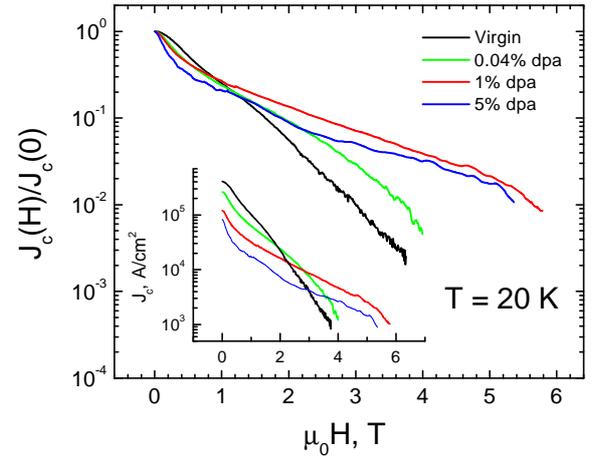

Fig. 3

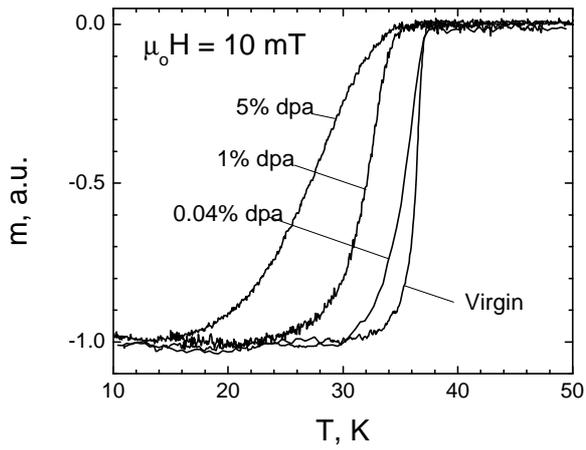

Fig. 2

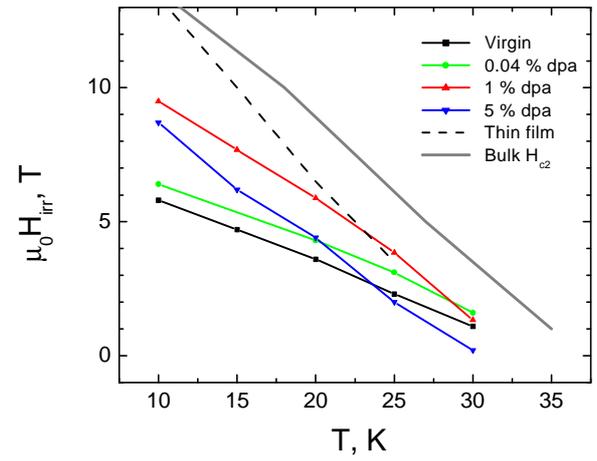

Fig. 4